# Thermoelectric Properties of $YBa_2Cu_3O_{7-\delta}$ – $La_{2/3}Ca_{1/3}MnO_3$ Superlattices

S. Heinze, H.-U. Habermeier, G. Cristiani, S. Blanco Canosa, M. Le Tacon,

and B. Keimer

Max Planck Institute for Solid State Research, Heisenbergstr. 1,

70569 Stuttgart, Germany

**Abstract**

We report measurements of the thermoelectric power and electrical resistivity of superlattices composed of the high-temperature superconductor $YBa_2Cu_3O_{7-\delta}$ (YBCO) and the metallic ferromagnet $La_{2/3}Ca_{1/3}MnO_3$ (LCMO) with individual layer thicknesses between 5 and 50 nm. Whereas YBCO and LCMO reference films prepared under the same conditions exhibit negative Seebeck coefficients, in excellent agreement with data on bulk compounds of identical composition, those of all superlattices are positive, regardless of the individual layer thickness. Having ruled out lattice strain and incomplete oxygenation, we attribute the observed sign reversal of the Seebeck coefficient to a long-range electronic reconstruction nucleated at the YBCO-LCMO interfaces.

The thermoelectric properties of metal oxides have been the subject of major research activity for the past two decades. The thermoelectric power is a measure of entropy rather than charge transport, and hence yields insights into the influence of electronic correlations on the physical properties complementary to those obtained from standard resistivity and Hall effect measurements.[1] In research on high-temperature superconductors, for instance, the thermoelectric power has served as a calibration standard for the doping level[2,3] and as a sensitive indicator of competing order and doping-induced Fermi surface reconstructions.[4] At the same time, metal-oxide compounds are being investigated for thermoelectric device applications in view of their compositional variability, chemical stability, and comparatively low cost.[5] A different avenue for fundamental[6] and applications-oriented[7] research on metal oxides has recently been established by advances in the synthesis of epitaxial heterostructures and multilayers with tailored lattice structure and electronic density of states, and the potential of such systems for research on thermoelectricity is just beginning to be recognized.[5,8-10] In particular, recent experiments on $SrTiO_3$-based heterostructures have demonstrated substantial modulations of the Seebeck coefficient in response to applied gate fields[9] and thickness variations of the $SrTiO_3$ layers.[10]

Aided by extensive knowledge of the thermoelectric properties of the bulk constituents, we have used a similar strategy to explore the electronic properties of superlattices of the high-temperature superconductor $YBa_2Cu_3O_{7-\delta}$ (YBCO) and the metallic ferromagnet $La_{2/3}Ca_{1/3}MnO_3$ (LCMO). YBCO-LCMO superlattices have served as a model system for the interplay between ferromagnetism and superconductivity,[11-13] charge transfer across oxide interfaces,[14,15] interfacial orbital[15] and magnetic[16-18] reconstructions, and electron-phonon interactions in multilayer systems.[19] While some of these phenomena appear to be restricted to the immediate vicinity of the interface, others (including a superconductivity-induced rearrangement of the magnetic structure of LCMO[17,18] and the hybridization of YBCO and LCMO phonon modes[19]) persist over length scales of tens of nanometers. Here we show that the thermoelectric response of YBCO-LCMO superlattices differs qualitatively from the one of its bulk constituents, and that this modification persists over a similarly wide spatial range.

YBCO-LCMO superlattices as well as YBCO and LCMO reference films were grown by pulsed laser deposition on (100)-oriented single-crystalline $SrTiO_3$ substrates. For comparison, an LCMO film was also grown on a (100)-oriented $(La_{0.27}Sr_{0.73})(Al_{0.65}Ta_{0.35})O_3$ (LSAT) substrate. All structures were deposited at 730 °C at an oxygen pressure of 0.5 mbar and annealed at 530 °C for 1 h in 1 bar oxygen. X-Ray diffraction analysis confirmed the phase purity of the c-axis oriented films (Fig. 1). All superlattices were fabricated with YBCO as the first layer and LCMO as the top layer. All samples were contacted with Cr/Au 20/200 nm pads on the LCMO top layer to reduce the thermal and electrical contact resistance compared to contacts directly fabricated on the sample surface by silver epoxy. The thermopower and resistivity were measured either with the Thermal Transport Option of a Quantum Design Physical Property Measurement System (PPMS) or an Ulvac ZEM-3 M8. The critical temperatures $T_c$ for superconductivity and $T_{Curie}$ for ferromagnetism were determined from the resistivity anomalies associated with these transitions (Fig. 2), following prior work.[11] Table I lists the transition temperatures resulting from these measurements, along with the instruments used to take the corresponding data and the layer thicknesses.

The oxygen content and strain state of the films and selected superlattices were determined by x-ray diffractometry. Analysis of the specular x-ray diffraction patterns at room temperature (Fig. 1) yielded the c-axis lattice parameter of YBCO, which is a sensitive measure of the oxygen content.[20] The values of 11.695(5) Å for the YBCO film and the superlattices YBCO-LCMO 1 and 2 (Fig. 2 and Table I) were found to be identical within the error, and consistent with the values of other superlattices prepared in an identical manner reported previously.[19] This value is in excellent agreement with the one reported for optimally doped YBCO single crystals ($\delta = 0.05$)[20] and with the results of Raman-scattering measurements of an apical-oxygen vibration whose frequency depends sensitively on the oxygen content.[19] The strain state of selected superlattice samples was characterized by extensive sets of reciprocal-space maps at temperatures between 10 and 300 K (see Supplementary Materials of Ref. 19), which showed that the in-plane lattice parameters do not exhibit any significant deviations from those of bulk YBCO and LCMO over the entire temperature range. This is expected because of the minute (< 0.3%) difference between them, and because of the large film thicknesses (100-300 nm) which ensure that the substrate-induced strain is fully relaxed. Neither strain imposed by the substrate nor mutual strain of the constituents therefore affect the physical properties in a major way.

Figure 2 shows the temperature dependent Seebeck coefficient, $S$, and resistivity, $\rho$, parallel to the substrate plane of two representative symmetric YBCO-LCMO superlattices with the same overall thickness (200 nm), but YBCO and LCMO thicknesses differing by a factor of two (13 and 26 nm, respectively), along with data on YBCO and LCMO reference films with the same overall thickness. The data on the reference films are in excellent agreement with prior data on bulk compounds of the same composition. In particular, the Seebeck coefficient of YBCO is very small and negative (< 3 µV/K) over the entire temperature range, as observed for optimally doped YBCO.[2-3,21-23] Since $S$ is positive and its magnitude is considerably larger for underdoped YBCO,[2-4,21-23] this observation implies that the oxygenation state of the film is close to the one corresponding to optical doping (δ ~ 0.05). This reasoning also applies to the YBCO-LCMO superlattices, which were prepared in an identical manner, thus confirming the conclusion of the x-ray experiments described above. Further, the close agreement of the Seebeck coefficient of the YBCO film with those of optimally doped YBCO underscores our conclusion that strain imposed by the $SrTiO_3$ substrate or extended defects nucleated by substrate imperfections do not affect the physical properties of our samples significantly.

The thermopower of the LCMO films is also negative and very small in the ferromagnetic metallic state at low temperature, but increases in amplitude upon heating to $T_{Curie}$ ~ 275 K and is weakly temperature dependent in the paramagnetic state (Fig. 2). This behavior again reproduces the one found in prior experiments on bulk LCMO, where it was attributed to interplay between entropy transport by localized-spin and itinerant-electron systems.[24-26] Films grown on $SrTiO_3$ and LSAT behave in a very similar manner, confirming our conclusion that substrate-induced strain or substrate-specific extended defects do not influence the thermoelectric properties in a major way.

In contrast to the constituent materials, the Seebeck coefficients of the superlattice samples are positive throughout the temperature range measured (Fig. 2). Clearly, a superposition according to Kirchhoff's law:

$$S \times \rho^{-1} \times d = \sum_i S_i \times \rho_i^{-1} \times d_i \qquad (1)$$

with $i$ = YBCO, LCMO cannot account for the observed sign reversal. This finding agrees with a prior report on YBCO-LCMO bilayers,[27] although in this work a corresponding YBCO reference film also showed positive $S$ with comparable magnitude, so that incomplete oxygenation could not be ruled out. The x-ray diffractometry data on our samples and the comparison with the YBCO reference film allow us to exclude this possible origin.

Since the Seebeck coefficients in superlattices with 13 and 26 nm thick individual layers are closely similar, we can also rule out a scenario in which the YBCO-LCMO interface is responsible for this effect, in contrast to the $SrTiO_3$-$GdTiO_3$ heterostructures already mentioned above, where the two-dimensional electron gas generated by the polarization discontinuity at the interface dominates the transport properties.[10] Indeed, prior work on the YBCO-LCMO system has shown that the interfacial region is more insulating than the constituent bulk materials,[14] probably as a consequence of charge transfer across the interface[14,15] and an orbital reconstruction[15] that reduces the conduction band width close to the interface. In agreement with this expectation, the thermopower of a superlattice with 5 nm thick YBCO layers (YBCO-LCMO 7 in Table 1) is significantly lower in magnitude (while its sign is still positive).

Table I reports the Seebeck coefficients of several other YBCO-LCMO superlattices at a temperature of 355 K, well above the critical temperatures for the superconducting and ferromagnetic phase transitions. Apart from the superlattice with the thinnest YBCO layers mentioned above, all of the values are in the range +10-20 µV/K, including superlattices with up to 50 nm thick YBCO layers (YBCO-LCMO 11 in Table I). This is surprising at first sight, because layers of this thickness are expected to behave in a bulk-like manner, but consistent with the recent observation of a modification of the electron-phonon interactions of YBCO-LCMO superlattices over a similarly long range.[19]

Since we ruled out oxygen deficiency and strain as major contributors to the long-range modification of the thermoelectric properties of the superlattices, we attribute this effect to an electronic reconstruction nucleated at the interfaces, but propagating well inside the layers. A related effect has recently been reported for $VO_2$-based heterostructures, where electrostatic charge accumulation at the surface was observed to drive a metal-insulator transition across a 70 nm thick film.[28] Our data on the YBCO-LCMO system could be explained by a charge-ordered state nucleated at the magnetically and orbitally reconstructed interfaces. This scenario is supported by the observation that bulk underdoped YBCO materials exhibit Seebeck coefficients of similar sign and magnitude as the ones we have observed in our superlattices.[2-4,21-23] The difference between these data and corresponding data on optimally doped YBCO (where *S* is negative and much smaller in magnitude) has been ascribed to a Fermi surface reconstruction due to a (real or incipient) charge-ordering instability in the underdoped regime.[4] Direct evidence of such an instability has recently been reported in underdoped bulk YBCO.[29] The search for direct evidence of charge order in YBCO-LCMO superlattices is an interesting subject of further investigation. Independent of the mechanism driving the phenomenon we have observed, our data suggests that phase control of correlated-electron systems may offer interesting perspectives for thermoelectric device applications.

**Acknowledgement**

The authors gratefully acknowledge N. Driza, G. Khaliullin, S.Hebert, and G. Logvenov for discussions and M. Schulz for technical support. Financial support was provided by the German Science Foundation under collaborative research grant

SFB/TRR 80 and by the European Commission under the project "SOPRANO" (Marie Curie actions grant no. PITNGA-2008-214040)..

**References**

1. K. Behnia, D. Jaccard, and J. Flouquet, J. Phys.: Condens. Matter **16**, 5187 (2004).
2. J. L. Tallon, C. Bernhard, H. Shaked, R. L. Hitterman, and J. D. Jorgensen, Phys. Rev. B **51**, 12 911(R) (1995).
3. J.L. Tallon, J.R. Cooper, P.S.I. P. N. de Silva, G. V. M. Williams, and J.W. Loram, Phys. Rev. Lett. **75**, 4114 (1995).
4. F. Laliberté, J. Chang, N. Doiron-Leyraud, E. Hassinger, R. Daou, M. Rondeau, B.J. Ramshaw, R. Liang, D.A. Bonn, W.N. Hardy, S. Pyon, T. Takayama, H. Takagi, I. Sheikin, L. Malone, C. Proust, K. Behnia, and L. Taillefer, Nature Comm. **2**, 243 (2011).
5. J. He, Y. Liu, and R. Funahashi, J. Mater. Res. **26**, 1762 (2011).
6. H.Y. Hwang, Y. Iwasa, M. Kawasaki, B. Keimer, N. Nagaosa, and Y. Tokura, Nature Mater. **11**, 103 (2012).
7. J. Mannhart and D. G. Schlom, Science **327**, 1607 (2010).
8. P. X. Zhang and H.-U. Habermeier, J. of Nanomaterials 329601 (2008).
9. H. Ohta, T. Mizuno, S. Zheng, T. Kato Y. Ikuhara, K. Abe, H. Kumomi, K. Nomura, and H. Hosono, Adv. Mater. **24**, 740 (2012).
10. T. A. Cain, S. B. Lee, P. Moetakef, L. Balents, S. Stemmer, and S. J. Allen, Appl. Phys. Lett. **100**, 161601 (2012).
11. H.-U. Habermeier, G. Cristiani, R.K. Kremer, O. Lebedev, and G. van Tendeloo, Physica C **364-365**, 298 (2001).

12. Z. Sefrioui, D. Arias, V. Peña, J. E. Villegas, M. Varela, P. Prieto, C. León, J. L. Martinez, and J. Santamaria, Phys. Rev. B **67**, 214511 (2003).

13. C. Visani, Z. Sefrioui, J. Tornos, C. Leon, J. Briatico, M. Bibes, A. Barthélémy, J. Santamaria, and J. E. Villegas, Nature Phys. **8**, 539 (2012).

14. T. Holden, H.-U. Habermeier, G. Cristiani, A. Golnik, A. Boris, A. Pimenov, J. Humlíček, O. I. Lebedev, G. Van Tendeloo, B. Keimer, and C. Bernhard, Phys. Rev. B **69**, 064505 (2004).

15. J. Chakhalian, J. W. Freeland, H.-U. Habermeier, G. Cristiani, G. Khaliullin, M. van Veenendaal, and B. Keimer, Science **318**, 1114 (2007).

16. J. Chakhalian, J.W. Freeland, G. Srajer, J. Strempfer, G. Khaliullin, J.C. Cezar, T. Charlton, R. Dalgliesh, C. Bernhard, G. Cristiani, H.-U. Habermeier, and B. Keimer, Nature Phys. **2**, 244 (2006).

17. J. Hoppler, J. Stahn, Ch. Niedermayer, V. K. Malik, H. Bouyanfif, A. J. Drew, M. Rössle, A. Buzdin, G. Cristiani, H.-U. Habermeier, B. Keimer, and C. Bernhard, Nature Mater. **8**, 315 (2009).

18. D. K. Satapathy M. A. Uribe-Laverde, I. Marozau, V. K. Malik, S. Das, Th. Wagner, C. Marcelot, J. Stahn, S. Brück, A. Rühm, S. Macke, T. Tietze, E. Goering, A. Frañó, J.-H. Kim, M. Wu, E. Benckiser, B. Keimer, A. Devishvili, B. P. Toperverg, M. Merz, P. Nagel, S. Schuppler, and C. Bernhard, Phys. Rev. Lett. **108**, 197201 (2012).

19. N. Driza, S. Blanco-Canosa, M. Bakr, S. Soltan, M. Khalid, L. Mustafa, K. Kawashima, G. Christiani, H.-U. Habermeier, G. Khaliullin, C. Ulrich, M. Le Tacon, and B. Keimer, Nature Mater. **11**, 675 (2012).

20. R. Liang, D. A. Bonn, and W. N. Hardy, Phys. Rev. B **73**, 180505(R) (2006).

21. P. J. Ouseph and M. R. O'Bryan, Phys. Rev. B **41**, 4123 (1990).

22. J. R. Cooper, S. D. Obertelli, A. Carrington, and J. W. Loram, Phys. Rev. B **44**, 12086 (1991).

23. J. W. Cochrane, G. J. Russell, and D. N. Matthews, Physica C **232**, 89 (1994).

24. M. F. Hundley and J. J. Neumeier, Phys. Rev. B **55**, 11511 (1997).

25. T. T. Palstra, A. P. Ramirez, S-W. Cheong, B. R. Zegarski, P. Schiffer, and J. Zaanen, Phys. Rev. B **56**, 5104 (1997).

26. P. Mandal, Phys. Rev. B **61**, 14675 (2000).

27. J. G. Lin, S. L. Cheng, C. R. Chang, and D. Y. Xing, J. Appl. Phys. **98**, 023910 (2005).

28. M. Nakano, K. Shibuya, D. Okuyama, T. Hatano, S. Ono, M. Kawasaki, Y. Iwasa, and Y. Tokura, Nature **487**, 459 (2012).

29. G. Ghiringhelli, M. Le Tacon, M. Minola, S. Blanco-Canosa, C. Mazzoli, N. B. Brookes, G. M. De Luca, A. Frano, D. G. Hawthorn, F. He, T. Loew, M. Moretti Sala, D. C. Peets, M. Salluzzo, E. Schierle, R. Sutarto, G. A. Sawatzky, E. Weschke, B. Keimer, and L. Braicovich, Science **337**, 821 (2012).

Table I. Properties of the samples investigated in this work. (ND: transition temperatures were not determined).

| Sample | Thickness [nm] | $T_C$ [K] | $T_{Curie}$ [K] | $S$ (355 K) [µV/K] | Instrument |
|---|---|---|---|---|---|
| YBCO | 200 | 86 | - | -0.8 ± 0.1 | PPMS |
| LCMO 1 (STO) | 200 | - | 275 | -15.4 ± 0.1 | PPMS |
| LCMO 2 (LSAT) | 200 | - | 260 | -12.3 ± 0.1 | PPMS |
| YBCO/LCMO 1 | 4x 26/26 | 68 | 260 | 12.2 ± 0.1 | PPMS |
| YBCO/LCMO 2 | 8x 13/13 | 59 | 245 | 14.2 ± 0.1 | PPMS |
| YBCO/LCMO 3 | 15x 10/10 | 45 | 230 | 18.2 ± 1.3 | ZEM |
| YBCO/LCMO 4 | 15x 10/10 | ND | ND | 19.2 ± 1.3 | ZEM |
| YBCO/LCMO 5 | 5x 10/10 | 62 | 240 | 11.2 ± 0.8 | ZEM |
| YBCO/LCMO 6 | 9x 6/6 | 34 | 230 | 15.9 ± 1.1 | ZEM |
| YBCO/LCMO 7 | 20x 5/10 | 35 | 235 | 3.5 ± 0.2 | ZEM |
| YBCO/LCMO 8 | 12x 15/10 | 48 | 225 | 21.5 ± 1.5 | ZEM |
| YBCO/LCMO 9 | 12x 15/10 | ND | ND | 16.0 ± 1.1 | ZEM |
| YBCO/LCMO 10 | 10x 20/10 | 60 | 220 | 17.0 ± 1.2 | ZEM |
| YBCO/LCMO 11 | 5x 50/10 | 82 | 230 | 13.5 ± 1.0 | ZEM |

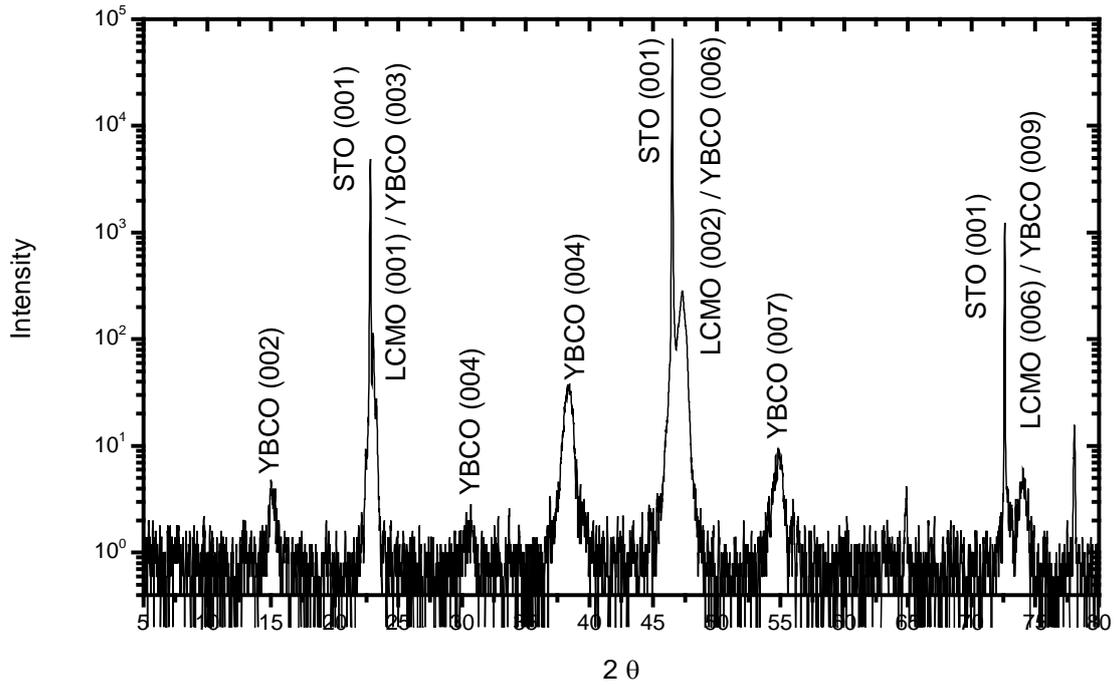

Fig.1. X-ray diffraction pattern of a symmetric superlattice with 13nm thick YBCO and LCMO layers (YBCO/LCMO 2 in Table I). The data were taken with Cu $K_\alpha$ radiation (wavelength 1.54 Å).

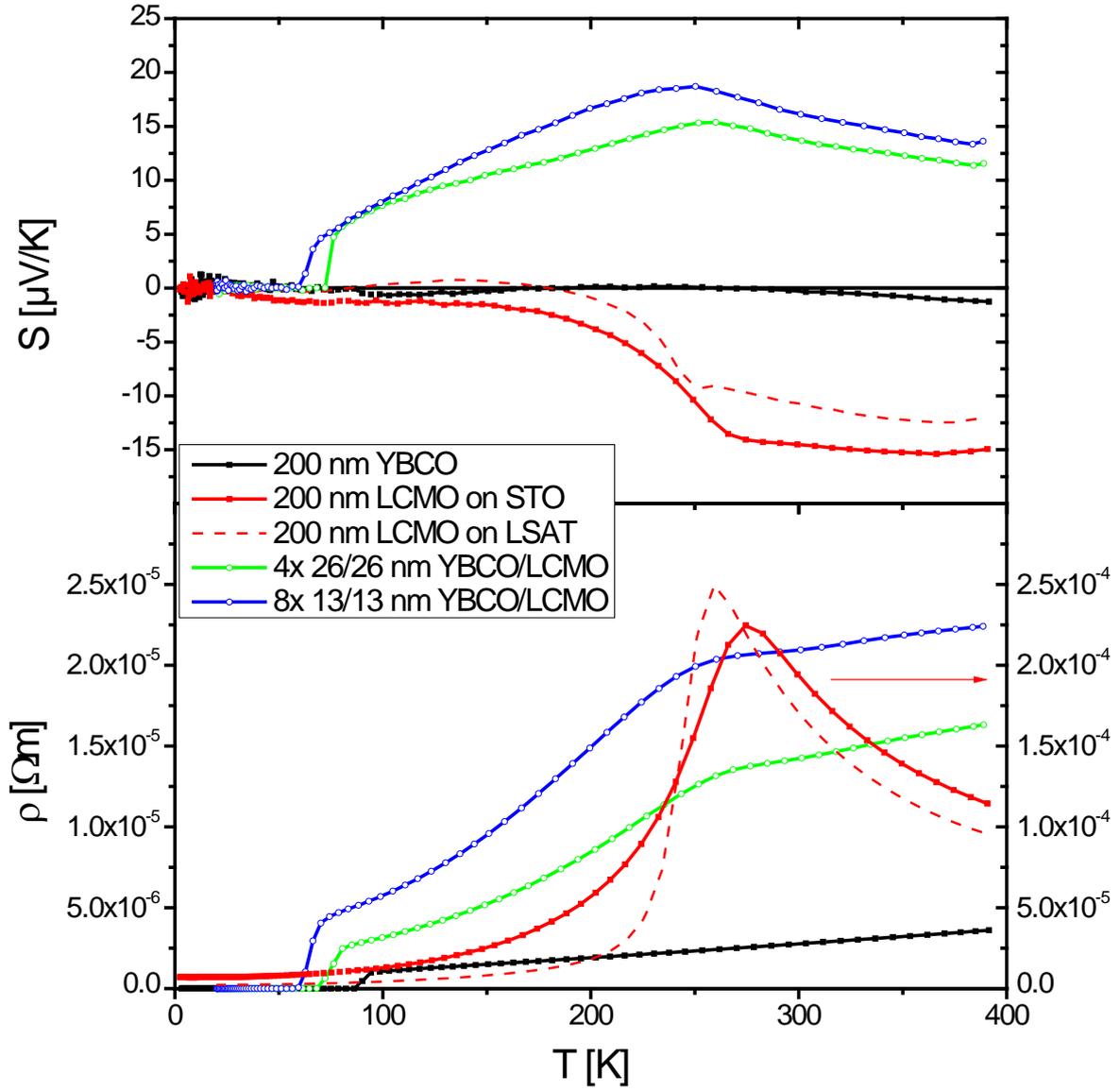

Fig. 2. Temperature (T) dependent Seebeck coefficient, *S*, and resistivity, *ρ*, for different samples, as stated in the legend.